\documentclass[prl,twocolumn,showpacs,preprintnumbers,amsmath,amssymb]{revtex4}
\usepackage{amsmath}
\usepackage{graphics}
\usepackage{epsf}
\usepackage{dcolumn}
\usepackage{bm}
\begin{document}
\title{Chalcogen-height dependent magnetic interactions and magnetic 
order switching in FeSe$_x$Te$_{1-x}$}
\author{Chang-Youn Moon and Hyoung Joon Choi}
\email[Email:\ ]{h.j.choi@yonsei.ac.kr}
\affiliation{Department of Physics and IPAP, Yonsei University, Seoul 120-749, Korea}

\date{\today}

\begin{abstract}
Magnetic properties of iron chalcogenide superconducting materials are
investigated using density functional calculations. We find the stability of magnetic
phases is very sensitive to the height of chalcogen species from the Fe
plane: while FeTe with optimized Te height has the double-stripe-type $(\pi,0)$
magnetic
ordering, the single-stripe-type $(\pi,\pi)$ ordering becomes the ground state phase
when Te height is lowered below a critical value by, e.g., Se doping. This behavior 
is understood by opposite Te-height dependences
of the superexchange interaction and a longer-range magnetic interaction
mediated by itinerant electrons. We also demonstrate a linear temperature dependence
of the macroscopic magnetic susceptibility in the single-stripe phase in contrast to
a constant behavior in the double-stripe phase. Our findings provide a comprehensive 
and unified view to understand the magnetism in FeSe$_x$Te$_{1-x}$ and iron pnictide
superconductors.
\end{abstract}


\maketitle
In a short period of time since the recent discoveries \cite{Kamihara2006,Kamihara2008,Takahashi},
the research of iron-based superconductors (SC) has been extended to a large variety of
materials. Among them, iron chalcogenides FeSe$_x$Te$_{1-x}$ are unique with their structural
simplicity that they are composed of only iron-based layers while maintaining the same nominal
Fe$^{2+}$ charge state as the iron pnictides. Unlike
most of iron SCs which have to be substantially doped to suppress the inherent
antiferromagnetism (AFM) and develop the superconductivity, undoped FeSe is not magnetically
ordered, superconducting with $T_c \sim 9 K$ \cite{Hsu,McQueen}, while
undoped FeTe is not superconducting but magnetically ordered. Very uniquely, FeTe has an AFM
ordering pattern, so called ``double stripe'' with $(\pi,0)$ ordering vector\cite{Li,Ma},
while all other iron-based superconducting materials exhibit ``single stripe'' type ordering
pattern with the $(\pi,\pi)$ ordering vector. 

Nature of magnetism in iron SCs 
has been a controversial issue between Fermi surface (FS) nesting
and local spin moment interactions. Several recent 
works suggest unified pictures based on local moment interactions including some role of
itinerant electrons but not in the way as in FS nesting scenario
\cite{Wu1,Wu2,Kou,Samolyuk,new1,new2,new3,LOFS,BFS}.
Since FeTe has the $(\pi,\pi)$ FS nesting similarly to other iron SCs \cite{Subedi}, 
the observed $(\pi,0)$ magnetic ordering in FeTe may indicate 
irrelevance of FS nesting mechanism at least for FeTe \cite{Ma}.
A recent first-principles study \cite{Han} suggests the emergence of $(\pi,0)$ FS
nesting assuming substantial electron doping by excess Fe, however, 
this doping effect was not observed in an angle-resolved photoemission experiment \cite{Xia}.

The magnetic stability of $(\pi,0)$ ordering over $(\pi,\pi)$ in FeTe can be effectively 
described by the nearest, second nearest, and third nearest neighbor exchange parameters, 
$J_1$, $J_2$, and $J_3$, respectively, with the condition $J_3 > J_2/2$ \cite{Ma}.
However, what makes such a condition hold uniquely for FeTe, neither FeSe nor iron
pnictides, is not uncovered. Furthermore, the absence of linear temperature ($T$) dependence
of magnetic susceptibility in FeTe \cite{Li,Xia}, in contrast to other iron SCs, is yet
to be understood.

In this paper, we perform the first-principles calculations to investigate
magnetic properties of iron chalcogenides. We find that Te height from the Fe 
plane is a key factor which determines AFM ordering patterns
in FeTe,
so that the ground state magnetic ordering changes from the $(\pi,0)$ with the 
optimized Te height to the $(\pi,\pi)$ patterns when Te height is lowered. We observe the 
same effect with FeSe, concluding that it is not chalcogen species but chalcogen 
position
that determines the magnetic ordering in FeSe$_x$Te$_{1-x}$,
and that this feature originates from different
chalcogen-height dependences of $J_1$, $J_2$, and $J_3$. 
Our calculated macroscopic magnetic susceptibilities
($\chi_M$) for $(\pi,\pi)$ and $(\pi,0)$ AFM orderings show linear and constant $T$ 
dependences, respectively, suggesting a clue to understand the puzzling 
$T$ dependences of $\chi_M$ measured in iron SCs.
Our results, altogether, provide a comprehensive view
on the magnetism in FeSe$_x$Te$_{1-x}$ and iron pnictide SCs.

Our first-principles calculations are based on the density-functional theory with
the generalized gradient approximation \cite{PBE} and the {\it ab-initio} 
norm-conserving pseudopotentials as implemented in SIESTA 
code \cite{SIESTA}. Semicore pseudopotentials are used for Fe, and
electronic wave functions are expanded with localized pseudoatomic orbitals (double zeta 
polarization
basis set), with the cutoff energy of 500 Ry for real space mesh. Brillouin zone integration is
performed with 6 $\times$ 6 $\times$ 6 k-point grid
using a $2a \times 2a \times c$ supercell which contains 8 Fe and 8 Te atoms, and
is commensurate with both the double- and single-stripes AFM ordering patterns.

\begin{figure}
\epsfxsize 3.5in
\centerline{\epsffile{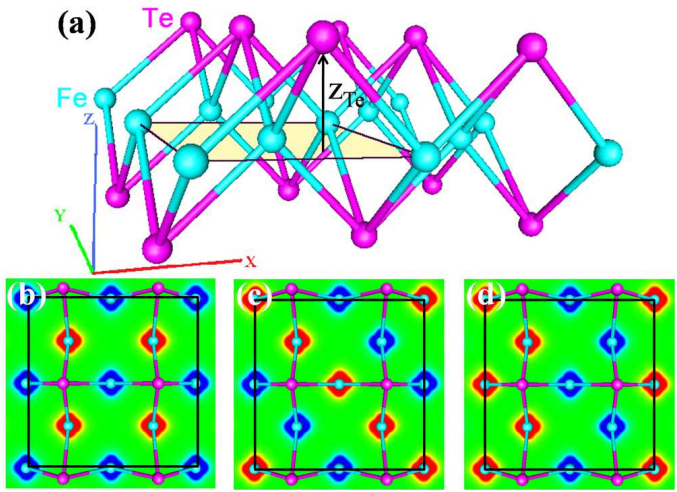}}
\caption{(Color online). (a) Atomic structure of FeTe where
$z_{Te}$ is defined with respect to the Fe plane marked by a shaded square.
Spin densities of (b) AFM1, (c) AFM2, and (d) AFM3 phases are displayed
on the $z=0$ plane, and the supercell is denoted by a black square in each figure.
}
\label{fig1}
\end{figure}

The atomic structure of FeTe is shown in Fig. 1(a). Fe atoms form a plane and Te atoms are
bonded to Fe atoms above and below the Fe plane, and the distance of a Te atom from
the Fe plane is defined as the Te height ($z_{Te}$) as shown in the figure. Structural 
optimization 
is performed with the unit-cell lattice constants fixed to the experimental high-temperature
tetragonal structure measured at 80 K,
$a=c=3.812$ {\AA} and $c=6.252$ {\AA} \cite{Li}, although the low temperature AFM phase
is associated with a monoclinic structure. This is for
simplicity in defining $z_{Te}$ and easy comparison with FeSe case. 
We have checked that this simplification does not alter our main conclusion.
We consider ferromagnetic (FM) phase and three different AFM phases to study 
the magnetic properties: `checkerboard' (Fig. 1(b)), `single stripe' (Fig. 1(c)), and 
`double stripe' 
(Fig. 1(d)) type orderings, which hereafter are referred to as AFM1, AFM2, and AFM3, 
respectively. 

To investigate effects of chalcogen height on magnetic orderings, we calculate
energetic stabilities of magnetic phases as a function of $z_{Te}$. 
Starting from the optimized $z_{Te}$ of 1.81 {\AA}, the total
energies of FM, AFM1, AFM2, and AFM3 phases are calculated with decreasing $z_{Te}$ as
shown in Fig. 2(a). At optimized $z_{Te}$, AFM3 is most stable 
in agreement with the experiment and the previous calculation \cite{Li,Ma},
and AFM2 is slightly higher in energy by 32 meV per Fe. Interestingly, FM phase is found to be
more stable than AFM1 phase in contrast to our previous experience on iron pnictides 
that FM phase relaxes to the non-magnetic (NM) state without a constraint on the magnetic moment 
and has higher energies than AFM phases even 
with a fixed spin moment (FSM) calculation. When $z_{Te}$ is decreased to 1.75 {\AA}, AFM3
is still more stable than AFM2 but the energy difference reduces. 
After $z_{Te}$ is further decreased
approximately below 1.71 {\AA} (see the vertical arrow in Fig. 2(a)),
finally AFM2 phase turns more stable than AFM3 phase, and the relative stability
of AFM2 over AFM3 phases is even enlarged as $z_{Te}$ is decreased further to 1.63 {\AA},
towards our optimized Se height 1.55 {\AA} of FeSe.
The Fe magnetic moment is decreasing along with $z_{Te}$, and the averaged values over 
the three AFM phases for each $z_{Te}$ are 2.94, 2.84, 2.72, and 2.58 $\mu_B$
for $z_{Te}$ = 1.81, 1.75, 1.69, and 1.63 {\AA}, respectively.
Except for the $z_{Te}$ =1.81 {\AA} case, we need to perform the FSM calculation
to obtain the FM phase,
in which the Fe moment is fixed to the average value of the three AFM phases with the 
same $z_{Te}$. 

\begin{figure}
\epsfxsize 3.5in
\centerline{\epsffile{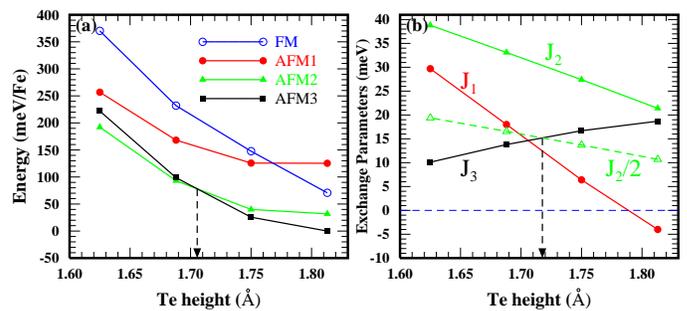}}
\caption{(Color online). (a) Total energy versus $z_{Te}$ calculated for
FM, AFM1, AFM2, and AFM3 phases. (b) Exchange parameters $J$'s are 
shown as functions of $z_{Te}$ which are evaluated with fixed Fe magnetic moment
of 2.95 $\mu_B$. Vertical arrows denote the critical $z_{Te}$'s where AFM2-AFM3 switch
takes place which are slightly different in (a) and (b) due to different treatment of
local magnetic moments (see text).
}
\label{fig2}
\end{figure}

Since AFM3 ordering in FeTe is apparently incompatible with the FS nesting 
scenario, we consider Heisenberg type local moment 
interactions, and estimate the three exchange parameters 
$J_1$, $J_2$, and $J_3$, as functions of $z_{Te}$, from the calculated total
energies of each magnetic phase and $z_{Te}$. Since $J$'s depend on the
magnitude of the magnetic moment in general \cite{Yildirim}, we perform the FSM
calculations, with the Fe spin moment being fixed to 2.95 $\mu_B$ (the value for AFM3 at 
$z_{Te}$=1.81 {\AA}), for all the magnetic phases and $z_{Te}$. The result is shown in
Fig. 2(b). 
At optimized $z_{Te}$ = 1.81 {\AA}, $J_2$ is largest, while $J_1$ is estimated to
be negative as the FM phase is more stable than AFM1 phase at this $z_{Te}$, and $J_3$ is
almost comparable with $J_2$. As $z_{Te}$ decreases, we see a clear tendency that $J_1$ 
and $J_2$ increase while
$J_3$ decreases. Around $z_{Te}$ = 1.72 \AA, $J_2/2$ starts to exceed $J_3$ which 
reflects that AFM2 phase becomes more stable than AFM3 from this point on.
We should note that this critical value of $z_{Te}$ is slightly different from the one in 
Fig. 2(a) where the magnetic moments are unrestricted, while they are fixed in
the estimation of $J$'s as mentioned above.

In a word, the switch of ground state magnetic phase between AFM2 and AFM3
with varying $z_{Te}$ is due to opposite $z_{Te}$ dependence between $J_2$ (along
with $J_1$) and $J_3$. The behavior of $J_1$ and $J_2$ can be understood
relatively easily if we recall their superexchange nature \cite{Yildirim}.
As Te atom gets close to the Fe plane, Fe-Te-Fe angle approaches 
$180^\circ$ which maximizes the overlap between Te $5p$ and Fe $3d$ 
orbital robes, and the superexchange interaction should get stronger.
As for $J_3$, a longer-range interaction is needed where isolated local
spin moments can be coupled via itinerant electrons around the Fermi energy
($E_F$), such as so-called RKKY interaction.

To understand $z_{Te}$ dependence of $J_3$, we consider how
a localized magnetic perturbation polarizes the surrounding initially 
spin-unpolarized
electron cloud. We apply a delta-like magnetic field on a Fe atom 
of the {\it FeTe system in the NM phase}, by 
adding a potential difference between spin up and down electrons at the
center of the Fe atom, and obtaining self-consistent electron spin density
in response to this external field, as displayed in Figs. 3(a) and (b).
By definition, this plot corresponds to $\chi_s(r,r'=0;\omega=0)$ when
the delta field is applied to the origin. Here we use $2 \times 2
\times 1$ enlarged supercell of our original supercell shown in
Fig. 1 to reduce the effect of other external fields from the periodic
images of the supercell. When Te is at the optimized height 1.81 {\AA}
(Fig. 3(a)),
the electron cloud is strongly spin-polarized by the delta field and show 
an alternating pattern, with a slow decay rate with distance from the origin.
In the meanwhile, when $z_{Te}$ is lowered down to 1.63 {\AA} in Fig. 3(b), 
the spin polarization decays much more rapidly
away from the origin. Hence, $\chi_s(r,r'=0;\omega=0)$ for the NM phase
becomes much more short-ranged with smaller $z_{Te}$, consistently
with the fact $J_3$ decreases with decreasing $z_{Te}$. 

\begin{figure}
\epsfxsize 3.5in
\epsffile{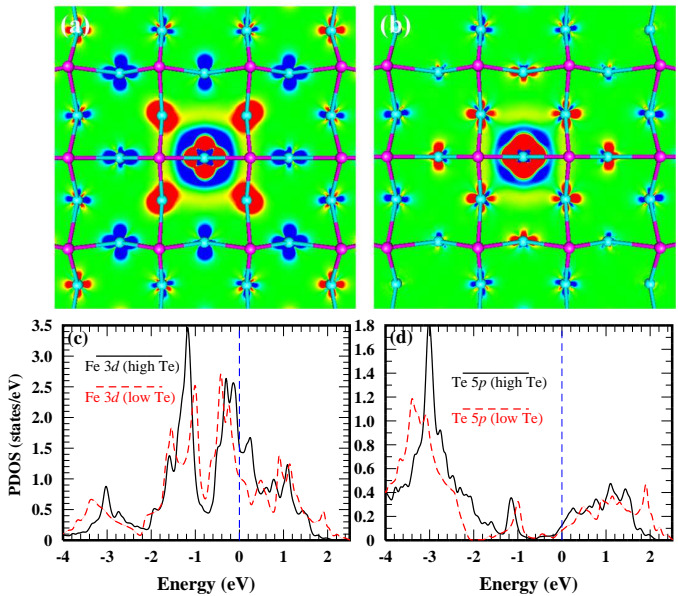}
\caption{(Color online). Magnetic susceptibility $\chi_s(r,r'=0;\omega=0)$ of
the NM phase is shown
for (a) $z_{Te}$=1.81 and (b) $z_{Te}=1.63$ {\AA}. Delta-like magnetic
field is applied at the center of the figure. PDOS in the NM phase is displayed
for (c) Fe $3d$ and (d) Te $5p$ states.
}         
\label{fig3}
\end{figure} 

The $z_{Te}$ dependence of $\chi_s$ can be understood
by analyzing the change in the electronic structure along with varying $z_{Te}$. 
Figure 3(c) represents the projected density of states (PDOS) on Fe $3d$,
where two peaks around -3 eV and $E_F$ have
$t_{2g}$ character. When Te is lowered, the $t_{2g}$ states couple
to Te $5p$ states more strongly so that the corresponding peaks around -3 eV
in Fig. 3(c) and (d) move down to the higher binding energy,
gaining more Fe $t_{2g}$ weight and losing Te $5p$ weight with an effective 
charge transfer from Te to Fe. Then the density of states (DOS) at $E_F$ is greatly reduced due to
the Fe $t_{2g}$ weight transfer to the states around -3 eV and overall charge
transfer to Fe states. This in turn results in generally reduced magnetic
susceptibility $\chi_s$, since it depends on the number of states 
around $E_F$ (which correspond to `itinerant electrons'). The higher DOS at 
$E_F$ for high $z_{Te}$ also explains the
stabilization of the FM phase at $z_{Te}$=1.81 {\AA} shown in Fig. 2, according
to the Stoner criterion.

\begin{figure}
\epsfxsize 3.5in
\centerline{\epsffile{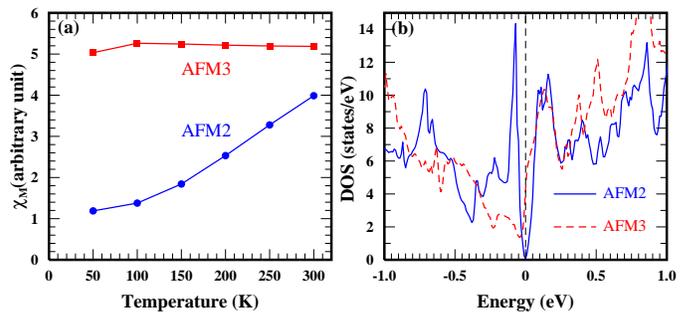}}
\caption{(Color online). (a) Macroscopic magnetic susceptibility $\chi_M$ 
versus electron temperature is shown 
for two AFM orderings. (b) DOS of FeTe for AFM2 and AFM3 phases.
}
\label{fig4}
\end{figure}

The Curie-Weiss-like behavior, i.e.,
the 1/$T$ dependence of magnetic susceptibility is a peculiar feature 
measured in FeTe \cite{Li,Xia} which undoubtedly shows the local nature of magnetism, 
while other iron-base SCs show linear-$T$ dependence. Motivated by the
fact the ground-state magnetic phase is AFM3 for FeTe and AFM2 for other
iron SCs, we evaluate the macroscopic magnetic susceptibility $\chi_M$
for different magnetic orderings, as a function
of electron temperature by applying a uniform magnetic field and
obtaining self-consistent electron spin density. The result is shown in Fig. 4(a).
The difference is obvious: $\chi_M$ depends linearly on $T$ for AFM2 while
it is almost constant for the AFM3 phase.
This difference results from the dramatic difference in DOS at $E_F$, shown
in Fig. 4(b). AFM2 phase has a large dip in DOS at $E_F$ because of the
band-repulsion between electron and hole bands which are separated by  
the $q=(\pi,\pi)$ vector in the Brillouin zone of the NM phase \cite{BFS}, 
while DOS at $E_F$ is large in AFM3 phase because
the perturbing potential by the magnetic ordering has only $q=(\pi,0)$
component which cannot couple the electron and hole FSs. Since 
the susceptibility
is larger when there is large number of states around $E_F$, the overall value
of $\chi_M$ is smaller for AFM2 phase. However, as the electron temperature
increases with the Fermi-Dirac distribution, more states become available to 
participate in the magnetic response and $\chi_M$ rises.

For $\chi_M$ in high-temperature paramagnetic phase, we
consider canonical ensemble of local spin moments, in which the probability of
each configuration is given by the Boltzmann factor associated mainly with the Heisenberg
Hamiltonian. When a uniform magnetic field is applied, $\chi_M$ has 1/$T$-like dependence
if the spin moments are frozen in each configuration. With itinerant electrons,
$\chi_M$ also has
contribution from field-induced spin-density change in each configuration averaged by the Boltzmann
factor, which is dominated by the ground state and low-energy configurations 
possibly composed of disordered domains
of the ground-state AFM phase \cite{Mazin2}. Then, the spin-density change in the 
ground-state AFM will 
contribute significantly to $\chi_M$ in the paramagnetic phase. Thus we suggest
$\chi_M$ in the paramagnetic phase has a linear-$T$ component for lower chalcogen
height, while the component is absent for higher chalcogen height.

We also have checked that the ground states in FeSe and LaFeAsO, which are AFM2 with 
optimized atomic positions, change to AFM3 when Se or As height is increased. 
This implies the magnetism in FeTe in fact is not qualitatively different 
from that of other iron-based SCs, and the peculiar $(\pi,0)$ ordering pattern is 
just a specific realization among many possible magnetic orderings determined
by the relative strength of local-moment exchange interactions $J$'s, which
can be controlled continuously by varying the chalcogen (and pnictogen) height, 
for example, by alloying such as FeSe$_x$Te$_{1-x}$. Although FeSe and other
alloys with FeTe are not experimentally reported to exhibit any magnetically 
ordered phase possibly due to the off-stoichiometry etc.,
our results suggest that AFM2 is the lowest-energy configuration 
in the canonical ensemble for paramagnetic FeSe$_x$Te$_{1-x}$
with $x$ over a critical value. This possibly results
in the linear-$T$ dependence of $\chi_M$ as demonstrated above. However, when Se
concentration is not enough, the chalcogen height is so high that AFM3 
is the lowest-energy configuration, and the Curie-Weiss-like 1/$T$ dependence 
of $\chi_M$ \cite{Li,Xia}, 
ordinary for the local moment magnetism, is observed in the paramagnetic phase.

Increasing Se in FeSe$_x$Te$_{1-x}$ is reported to suppress
$(\pi+\delta,0)$ spin fluctuation and enhance $(\pi,\pi)$ fluctuation with 
the superconductivity
\cite{Nakayama}. Our result shows that this might be related to the continuous
change of the relative energy between AFM3 and AFM2 orderings with increasing 
$x$. When $x$ is small and chalcogen height is large,
the AFM3 state and $(\pi,0)$ spin fluctuation has greater probability
in the canonical ensemble for the paramagnetic
phase, weakening the interaction between electron and hole FSs  
separated by $q=(\pi,\pi)$. As $x$ increases over the critical value, 
the AFM2 state and $(\pi,\pi)$ spin fluctuation start to take over,
which coincides with the superconductivity observed in FeSe$_x$Te$_{1-x}$.
Therefore, our result shows that the superconductivity in iron chalcogenides
may be related to $(\pi,\pi)$ spin ordering and fluctuations as likely as in 
iron pnictides. Our findings further suggest that the superconducting state might 
be realized in FeTe by controlling the Te height by, such as,
applying a biaxial strain.

In conclusion, we show that the unique magnetic ordering in FeTe is due to
the relatively high chalcogen height compared with FeSe as the underlying 
magnetic interactions depend on the chalcogen height in different ways. 
This behavior is found to apply for other iron chalcogenide and iron
pnictides, hence the magnetism in iron-based SCs including FeTe can
be understood with a same unified mechanism. Temperature dependence of 
magnetic susceptibility evaluated for the AFM2 and AFM3 phases sheds light
on the puzzling experimental results observed for FeSe$_x$Te$_{1-x}$.

\begin{acknowledgments}
This work was supported by NRF of Korea (Grant No. 2009-0081204) and
KISTI Supercomputing Center (Project No. KSC-2008-S02-0004). 

\end{acknowledgments}


\end{document}